\documentclass[a4paper,12pt]{article}
\usepackage{graphicx}
\usepackage{amsmath,amssymb} 
\usepackage{amsfonts}
\usepackage{natbib}
\usepackage[footnotesize]{subfigure}
\usepackage{color}
\usepackage{booktabs}
\usepackage{ucs}
\usepackage[T1]{fontenc}
\usepackage{times}
\usepackage{authblk}
\usepackage{array}
\usepackage{tabularx}
\usepackage{arydshln}
\usepackage{pdflscape}
\usepackage{tikz}
\usepackage{bbold}
\usepackage[margin=2.5cm]{geometry}
\usepackage[bf, small]{caption2}

\newcommand{\real}{\mathbb{R}}
\newcommand{\dd}{\textup{d}}

\title{Shape from Texture using Locally Scaled Point Processes}

\author{Eva-Maria Didden}
\affil{\small \em Institute of Applied Mathematics, Heidelberg University, Germany}
\author{Thordis L. Thorarinsdottir and Alex Lenkoski}
\affil{\small \em Norwegian Computing Centre, Oslo, Norway}
\author{Christoph Schn\"orr}
\affil{\small \em Image \& Pattern Analysis Group, Heidelberg University, Germany}
\date{\vspace{-3ex}}
\begin{document}
\maketitle
\begin{abstract}
\noindent
Shape from texture refers to the extraction of 3D information from 2D images with irregular texture.  
This paper introduces a statistical framework to learn shape from texture where convex texture elements in a 2D 
image are represented through a point process.  In a first step, the 2D image is preprocessed to generate a 
probability map corresponding to an estimate of the unnormalized intensity of the latent point process underlying 
the texture elements.  The latent point process is subsequently inferred from the probability map in a 
non-parametric, model free manner. Finally, the 3D information is extracted from the point pattern by applying a 
locally scaled point process model where the local scaling function represents the deformation caused by the 
projection of a 3D surface onto a 2D image. 

\vspace{0.4cm}


\end{abstract}

\section{Introduction}

\noindent
Natural images contain a variety of perceptual information enabling the viewer to infer the three-dimensional 
shapes of objects and surfaces \citep{Tuceryan1998}.  \cite{Stevens1980} observed that surface geometry 
mainly has three effects on the appearance of texture in images: foreshortening and scaling of texture elements, 
and a change in their density.  \cite{Gibson1950} proposed the slant, the angle between a normal to the surface
 and a normal to the image plane, as a measure for surface orientation. Stevens amended this by introducing the 
tilt, the angle between the surface normal's projection onto the image plane and a fixed coordinate axis in the 
image plane. In this paper, we will directly infer the surface normal from a single image taken under standard 
perspective projection.

Statistical procedures for estimating surface orientation often make strong assumptions on the regularity of texture.
\cite{Witkin1981} assumes observed edge directions provide the necessary information, while 
\cite{Blostein1989} consider circular texture elements with uniform intensity. \cite{Blake1990} consider the
bias of the orientation of line elements isotropically oriented on a 3D plane, induced by the plane's orientation under
orthographic projection, along with a computational approach related to Kanatani's texture moments 
\citep{Kanatani1989}.

\cite{Malik1997} locally estimate ``texture distortion'' in terms of an affine transformation of adjacent image 
patches.  The strong homogeneity assumption underlying this approach has been relaxed by \cite{Clerc2002}, 
to a condition that is difficult to verify in practice. \cite{Forsyth2006} eliminates assumptions on the non-local 
structure of textures (like homogeneity) altogether and aims to estimate shape from the deformation of individual 
texture elements.  \cite{Loh2005} criticize prior work due to the restrictive assumptions related to homogeneity, 
isotropy, stationarity or orthographic projection, and claim to devise a shape-from-texture approach in the most 
general form.  Their work, however, also relies on estimating the deformation of single texture elements, similar to
\cite{Forsyth2006}.

We propose a general framework for inferring shape from near regular textures, as defined by 
\cite{SymmetrySurvey-10}, by applying the locally scaled point process model of \cite{Hahn}.  This framework
enables the simultaneous representation of local variability and global regularity in the spatial arrangement of 
texture elements which are thought of as a marked point process.  We preprocess the image to obtain a probability 
map representing an unnormalized intensity estimate for the underlying point process, subsequently apply a
non-parametric framework to infer the point locations and based on the resulting point pattern, learn the 
parameters of a locally scaled point process model to obtain a compact description of 3D image attributes. 

Point process models have previously been applied in image analysis applications where the goal is the detection of 
texture elements, see e.g. \cite{Lafarge2010} and references therein.  These approaches usually apply a 
marked point process framework, with marks describing the texture elements.  Such set-ups rely on a good 
geometric description of individual texture elements, limiting the class of feasible textures.  As our goal is not the 
detection of individual texture elements but the extraction of 3D information, we omit the modeling of each texture 
element and infer the latent point locations in a model free manner.  Thus, our sole assumption regarding texture 
element shape is approximate convexity which offers considerable flexibility. 

The remainder of the paper is organized as follows. The next section contains preliminaries on image geometry 
followed by the method section describing the image preprocessing, the point pattern detection and the point process
inference framework. We then present results for both simulated and real images with near regular textures. Finally, 
the paper closes with a short discussion section. 

\section{Preliminaries}

\begin{figure}[t]
\centerline{
\includegraphics[width=0.4\textwidth]{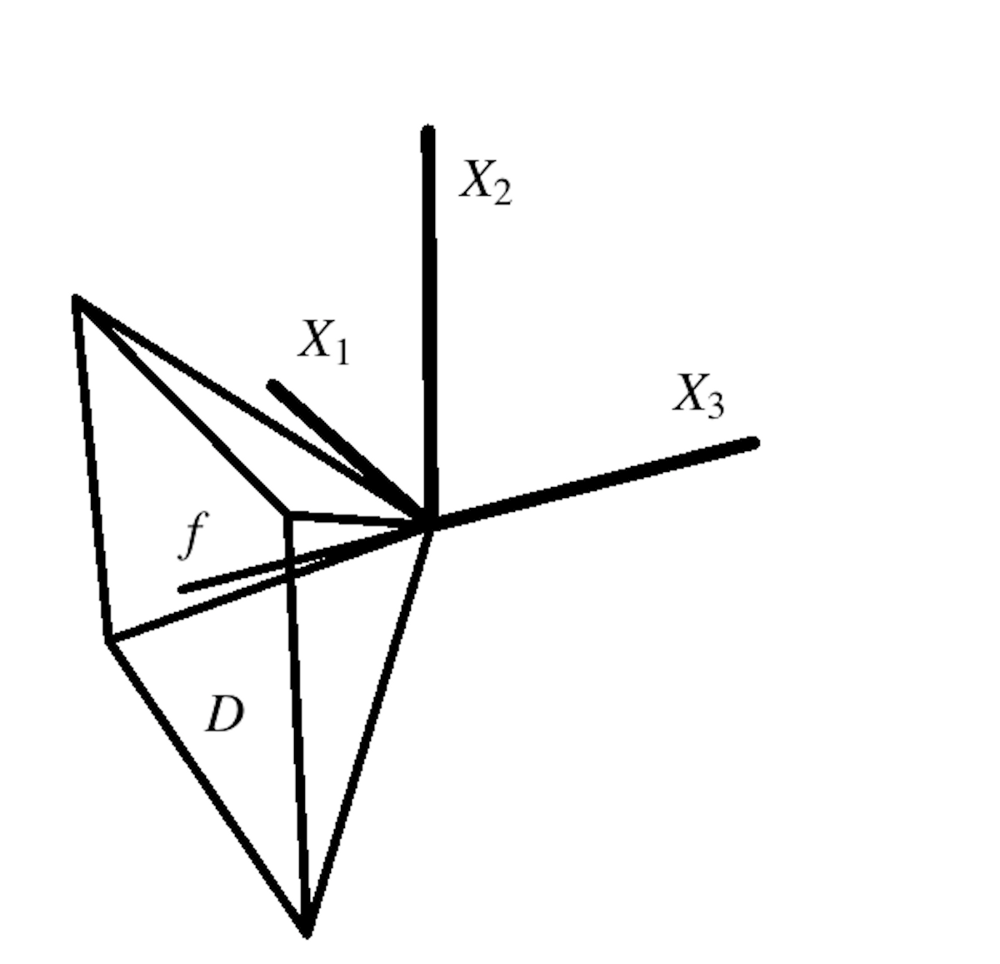}
}
\caption{
The camera with focal length $f$ is oriented towards the negative $X_{3}$-halfspace. The scaled visible image 
domain is $D = [-a/2,a/2] \times [-1/2,1/2]$.  Given the field of view in terms of an angle $\phi_{c}$, we have 
$f = \frac{a/2}{\tan(\phi_{c}/2)}$.
}
\label{fig:Imaging-Setup}
\end{figure}

\noindent
Let
\begin{equation} \label{eq:def-P}
 P = \{ X \in \real^{3} \colon \langle \delta, X \rangle + h = 0 \},
\end{equation}
with $\|\delta\|=1$ and  $\langle \delta, X \rangle < 0$,
denote a 3D plane with unknown unit normal $\delta$ and distance $h$ from the origin. We assume $\delta$ to be 
oriented towards the camera, forming obtuse angles $\langle \delta, X \rangle < 0$ with projection rays $X$.  The 
world coordinates $X = (X_{1},X_{2},X_{3})^{\top}$ and image coordinates $x = (x_{1},x_{2})^{\top}$ are 
aligned as shown in Fig. \ref{fig:Imaging-Setup}.  Here, we denote the image domain by $D$ and assume the image 
to be scaled to have fixed area, $|D| = a$. 

We consider the basic pinhole camera \citep{Hartley-Zisserman-00} and among the internal parameters, we 
only look at the focal length $f > 0$ which depends on the field of view, see Fig.~\ref{fig:Imaging-Setup}. As 
usual, we identify image points and rays of the projective plane through
\begin{equation} \label{eq:def-X-x}
 X = (x_{1}, x_{2}, -f )^{\top} \ .
\end{equation}
An image point $X$ given by \eqref{eq:def-X-x} meets $P$ in $\lambda X$ with 
\begin{equation}
\lambda = -\frac{h}{\langle \delta, X \rangle}, \quad \lambda > 0.
\end{equation}
It follows that a point $X_P$ in $P$ is related to the image point $X$ through
\begin{equation}
X_{P} = X_{P}(x_{1},x_{2}) = -\frac{h}{\langle \delta, X \rangle} X.
\end{equation}

\begin{figure}[ht!]
\centering
\subfigure[$\delta\hspace{-.1cm}=\hspace{-.1cm}(\frac{1}{\sqrt{2}}, 0, \frac{1}{\sqrt{2}})^\top$]{\includegraphics[width=3.5cm, height=3.5cm]{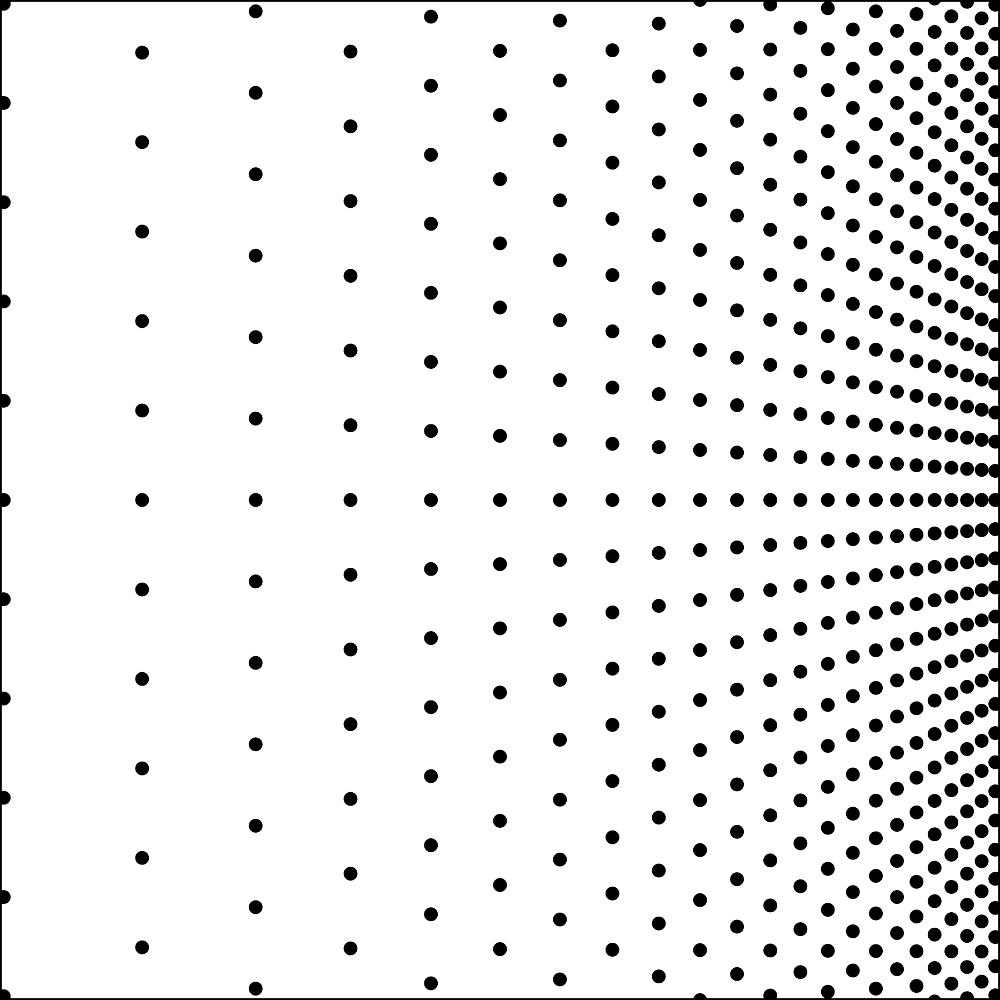}}
\hspace{0.5cm}
\subfigure[$\delta\hspace{-.1cm}=\hspace{-.1cm}(\frac{1}{2\sqrt{2}},\frac{1}{2\sqrt{2}},\frac{\sqrt{3}}{2})^\top$]{\includegraphics[width=3.5cm, height=3.5cm]{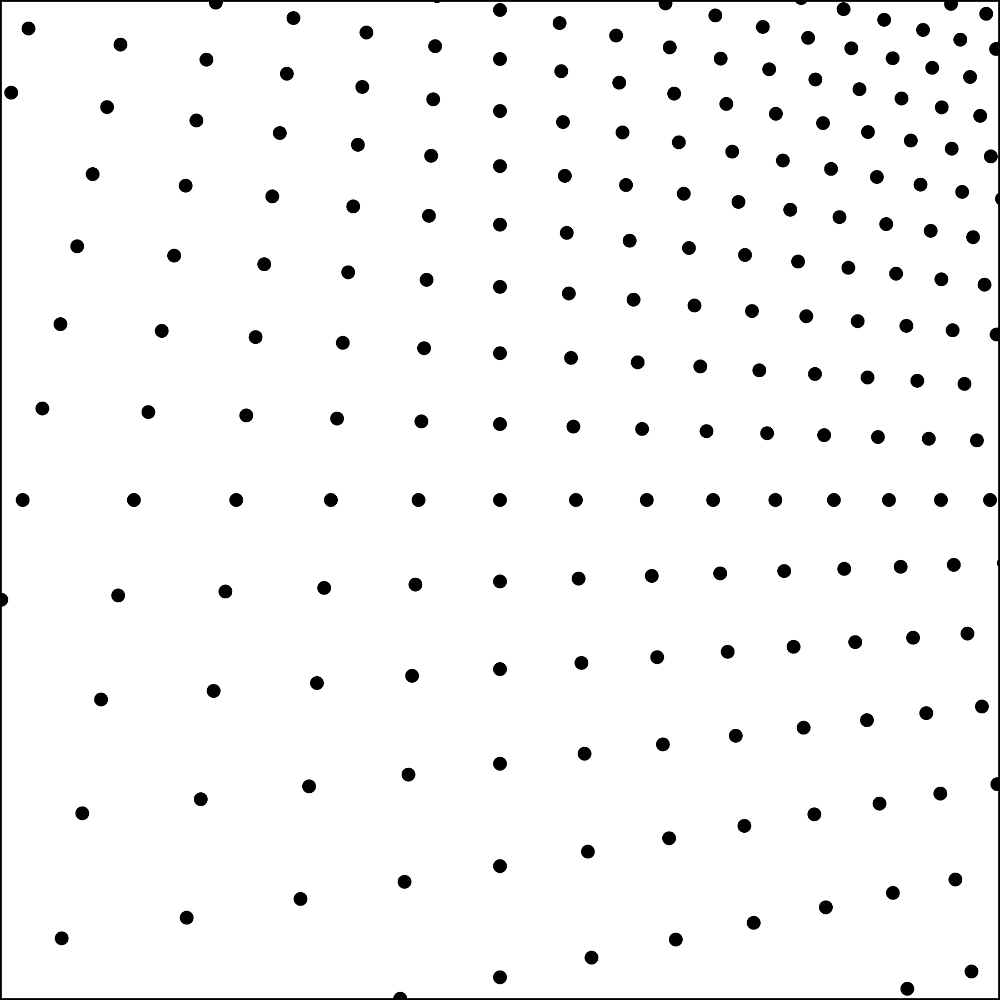}}
 \caption{Mappings of regular homogeneous point patterns in $\real^3$ onto a 2D-plane. The simulations are 
based on the parameters $D=[-1/2, 1/2]\times[-1/2, 1/2]$, $h=20$ and $\phi_c = 27^{\circ}$ ($f = 0.98$).}
\label{fig:regular points}
\end{figure}

A homogeneous texture covering $P$ induces an inhomogeneous texture on the two-dimensional image plane with density given by 
the surface element
\begin{align} \label{eq:XP-density}
 dX_{P} & = \|\partial_{x_{1}} X_{P} \times \partial_{x_{2}} X_{P}\|
 \lambda_{2}(\dd x) \nonumber \\
 & = -\frac{h^{2} \, f}{\langle \delta, X \rangle^{3}} \lambda_{2}(\dd x),
\end{align}
where $\lambda_2$ denotes the two-dimensional Lebesgue measure.  Taking, for instance, the fronto-parallel plane 
$\delta = (0,0,1)^{\top}$ results by \eqref{eq:def-X-x} merely in the constant scale factor $(h/f)^{2}$, 
i.e.~the homogeneous density $(h/f)^{2} \lambda_{2}(\dd x)$.  However, for arbitrary orientation $\delta$, this 
factor depends on $X$, as illustrated in Fig.~\ref{fig:regular points}.  Eqn.~\eqref{eq:XP-density} then 
quantifies perspective foreshortening and inhomogeneity of the texture, respectively, as observed in the image, 
and mathematically represents the visually apparent texture gradient. 

\section{Methods}
\label{sec: methods}

\noindent
In a first step, we apply image preprocessing that generates a probability map 
$Y = \{Y(x) \, : \, x \in D, \, 0 \leq Y(x) \leq 1 \}$ representing the spatial arrangement of texture elements 
in the image. To this end, two 
elementary techniques are locally applied: Boundary detection and the corresponding distance transform. The 
former step entails either gradient magnitude computation using small-scale derivative-of-Gaussian filters 
\citep{Canny1986} or, for texture elements with less regular appearance, the earth-mover's distance 
\citep{Pele2009} between local histograms. Inspecting in turn the histogram of the resulting soft-indicator 
function for boundaries enables one to determine a threshold and apply the distance transform. 

In our framework, the texture elements are regarded as a realization of a marked point process where the 
underlying point pattern is latent.  
The value of the probability map $Y(x)$ in $x \in D$ denotes the probability 
that one of the latent points is located in $x$.  To recover the latent point pattern based on the information in $Y$, 
we first search for local maxima in $Y$.  That is, for some $k_1 > 0$, let 
$W_x = [x_1 - k_1, x_1 + k_1] \times [x_2 - k_1, x_2 + k_1]$ and  set
\begin{equation}\label{eq:local maxima}
\Phi = \{ x \in D \, : \, W_x \subset D, \, Y(x) = \max_{z \in W_x} Y(z) \}.  
\end{equation}
We then define a neighbourhood relation on $\Phi$ by setting $x^1 \sim x^2$ if
\begin{equation}\label{eq:neighbourhood relation}
\min_{z \in [x^1,x^2]} Y(z) \, \geq \, k_2 \max\{Y(x^1), Y(x^2)\}, 
\end{equation}
where $x^1,x^2 \in \Phi$, $[x^1, x^2]$ denotes the line from $x^1$ to $x^2$ and $k_2$ is a constant with 
$0 < k_2 < 1$.   We may now write $\Phi$ as a union of disjoint neighbourhood components, 
$\Phi = \cup_{i=1,\ldots,n} \, C_i$, where each $x \in C_i$ is neighbour with at least one point in 
$C_i \setminus x$.   Under the assumption that the texture elements are close to convex, two points $x^1$ and 
$x^2$ in $\Phi$ are neighbours if and only if they likely fall 
within the same texture element. Hence, we estimate the latent point process $\Psi$ as  
\begin{equation}\label{eq:latent points}
\Psi = \{x^1,\ldots,x^n \, : \, Y(x^i) = \max_{z \in C_i} Y(z) \}. 
\end{equation}

Formally, a point process can be described as a random counting measure $N(\cdot)$, where $N(A)$ is the number 
of events in $A$ for a Borel set $A$ of the relevant state space, in our context the image domain $D$. The intensity 
measure of the point process is given by $\Lambda(A) = \mathbb{E} N(A)$ and the associated intensity function is
\begin{equation}\label{eq:intensity}
\alpha (x) = \lim_{| \dd x| \rightarrow 0} \frac{\mathbb{E} N(\dd x )}{| \dd x|}.
\end{equation}
For a homogeneous point process, it holds that $\alpha(x) = \beta$ for some $\beta > 0$, while for an 
inhomogeneous point process where the inhomogeneity stems from local scaling \citep{Hahn} we obtain 
\begin{equation}\label{eq:scaled intensity}
\alpha (x) = \beta c_\eta^{-2} (x),
\end{equation}
for some scaling function $c_{\eta} : \real^2 \rightarrow \real_+$ with parameters $\eta$.  The scaling function 
$c_{\eta}$ acts as a local deformation in that it locally affects distances and areas.  More precisely, 
$\nu_c^d(A) = \int_A c_\eta(x)^{-d} \nu^d (\dd x)$, where $v^d$ denotes the $d$-dimensional volume measure 
and $\nu_c^d$ its scaled version for $d= 1,2$.  

\begin{figure}        
\centering
\subfigure[$\eta = (-1,0)^\top$]{
  \includegraphics[width=0.21\textwidth]{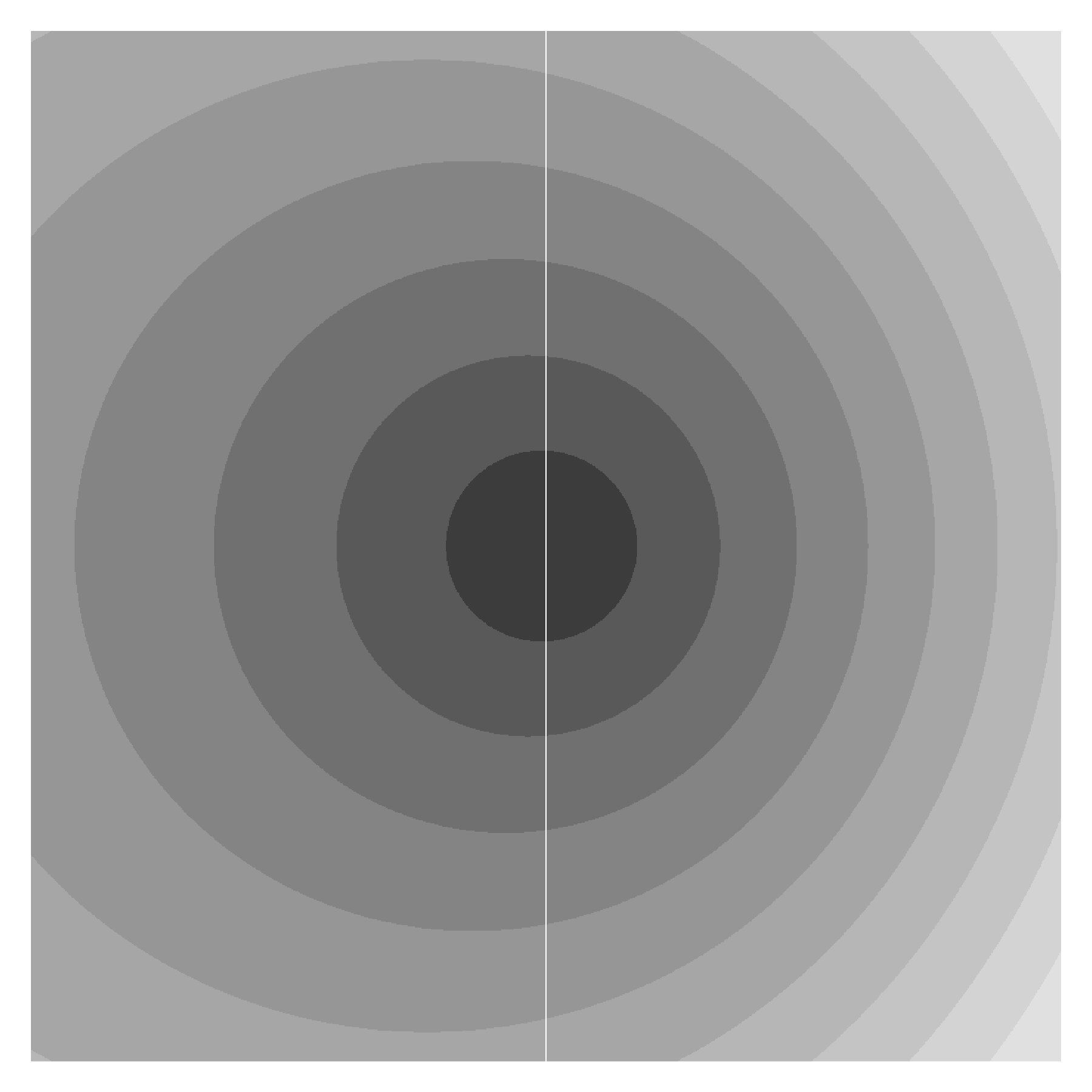}
}
\subfigure[$\eta = (-1,-1)^\top$]{
  \includegraphics[width=0.21\textwidth]{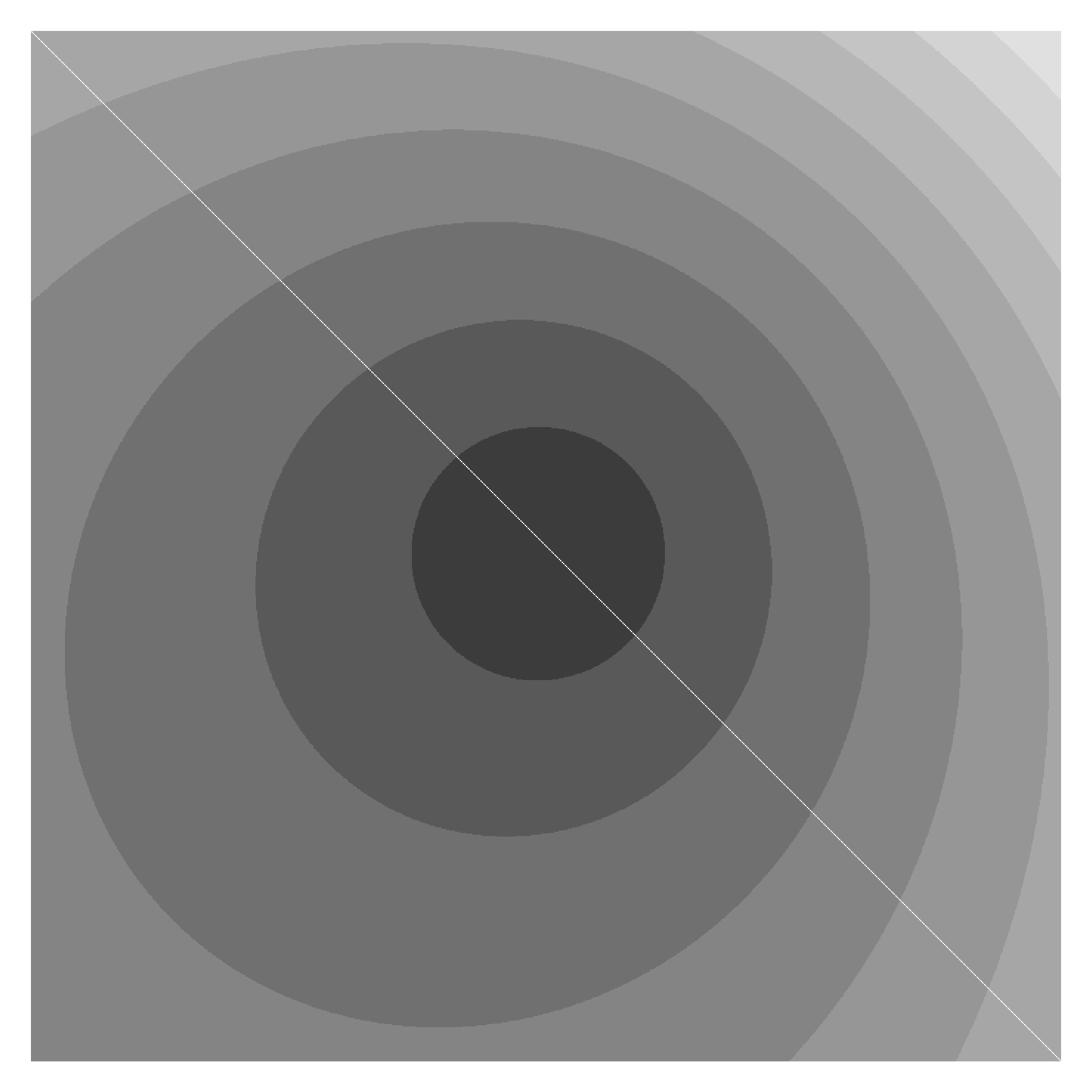}
}
\caption{Examples of distances from the point $(0, 0)$ within the observation window $D= [-1/2,1/2] \times [-1/2,1/2]$, 
under exponential scaling assumptions due to \eqref{eq:exponential scaling}. 
Darker shades of gray indicate smaller distances.}\label{fig:exponential scaling}
\end{figure}

For identifiability reasons, \cite{Prokesova} propose normalizing $c_{\eta}$ to conserve the total area of the 
state space.  That is, they define the normalizing constant of the scaling function such that 
\begin{equation}\label{eq:conservation area}
 \lambda_2(D) = \int_D c_{\eta}^{-2}(x) \lambda_2(\dd x).
\end{equation}
\cite{Hahn} and \cite{Prokesova} specifically consider the exponential scaling function with 
$c_{\eta}(x) \propto \exp(\eta^{\top} x)$.  This scaling function is particularly attractive in that locally scaled 
distances can be calculated explicitely, 
\begin{equation}
\label{eq:exponential scaling}
d_c(x^i,x^j) = d(x^i, x^j) \Big| \frac{c_{\eta}^{-1}(x^i) - c_{\eta}^{-1}(x^j)}{\eta^T (x^j - x^i)} \Big| \ ,
\end{equation}
for any $x^i, x^j \in D$ where $d(\cdot,\cdot)$ denotes the Euclidean distance and $d_c(\cdot, \cdot)$ its scaled 
version.  Examples of exponentially scaled distances are given in Fig.~\ref{fig:exponential scaling}.  

Here, we employ the density in \eqref{eq:XP-density} as a scaling function where we choose spherical coordinates
\begin{align}\label{eq:delta}
 \delta & = \delta(\eta_1,\eta_2) \\
 & = (\sin\eta_1 \cos\eta_2, \sin\eta_1 \sin\eta_2, \cos\eta_1)^{\top}, \nonumber
\end{align}
with  $\eta_1 \in [0,u]$ and  $\eta_2 \in [0,2 \pi]$.  The upper limit $u$ restricting the range of the scaling parameter $\eta_1$ 
ensures that $\langle \delta, X \rangle < 0$ and
therefore depends on the focal length $f$ as well as on the size and location of the observation window $D$.  As suggested by \cite{Prokesova}, we normalize the scaling function 
such that \eqref{eq:conservation area} holds.  That is, we solve
\begin{equation}
 |D| = a = \int_{D} \gamma (\delta, h, f) dX_{P}.  
\end{equation}
It follows that 
\begin{align}\label{eq:normalizing constant} 
\gamma (\delta, h,f) \, = \, & \frac{1}{16 h^2 f^2 \delta_{3}}
(a \delta_{1}-2 f \delta_{3}-\delta_{2}) \nonumber \\ 
& \quad \times (a \delta_{1}-2 f \delta_{3}+ \delta_{2}) \nonumber\\
& \quad \times (a \delta_{1}+2 f \delta_{3}- \delta_{2}) \nonumber \\
& \quad \times (a \delta_{1}+2 f \delta_{3}+ \delta_{2}) \ . \nonumber
\end{align}
A more general result for $D = [a_1, a_1] \times [b_1, b_2]$ is given in the Appendix.

\begin{figure}      
\centering
\subfigure[$\eta = (45^{\circ}, 0^{\circ})^\top$]{
  \includegraphics[width=0.21\textwidth]{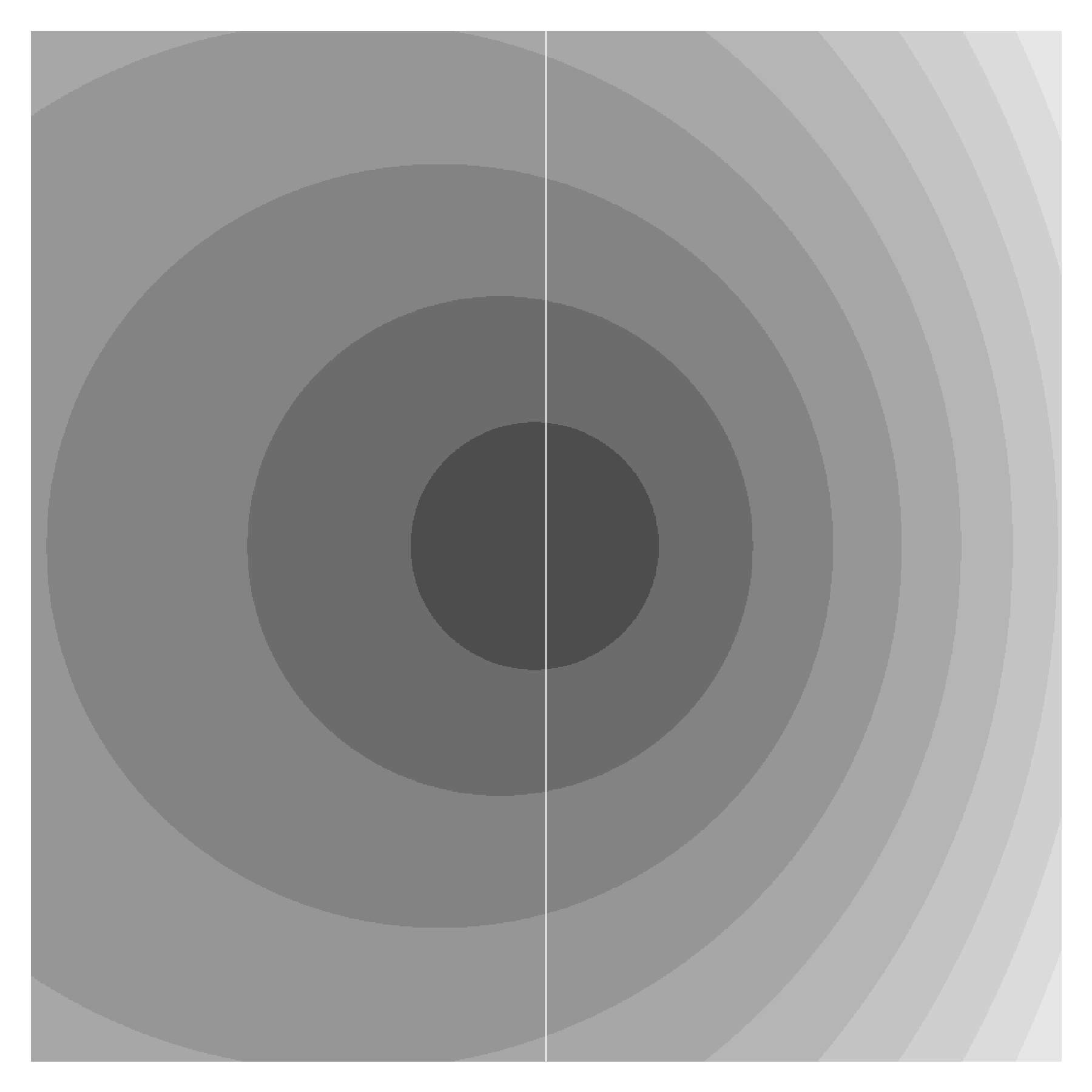}
}
\subfigure[$\eta = (30^{\circ}, 45^{\circ})^\top$]{
  \includegraphics[width=0.21\textwidth]{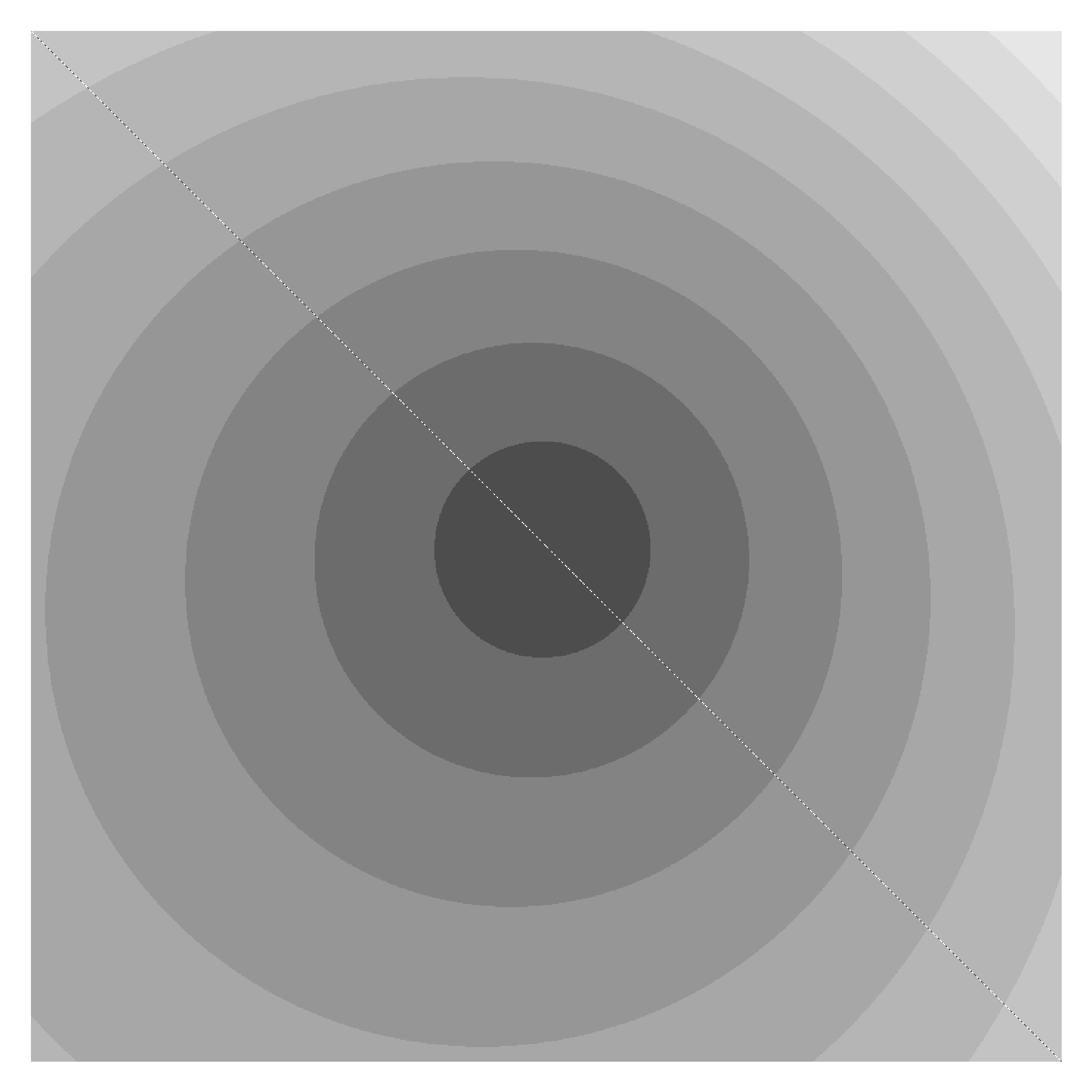}
}
\caption{Examples of distances from the point $(0, 0)$ within the observation window $D=[-1/2, 1/2] \times [-1/2, 1/2]$, 
under scaling assumptions due to \eqref{eq:scaled distances}. 
Darker shades of gray indicate smaller distances.}\label{fig:our scaling}
\end{figure}

Under the model in \eqref{eq:XP-density}, the intensity function in \eqref{eq:scaled intensity} becomes 
\begin{equation}\label{eq:model intensity}
 \alpha(x) = \beta \frac{\gamma\big(\delta(\eta_1,\eta_2),h,f\big) h^{2}\,f}{\big|\langle \delta(\eta_1,\eta_2), X \rangle\big|^{3}},
\end{equation}
with $X = (x_1,x_2, -f)^{\top}$ as in \eqref{eq:def-X-x}. As a byproduct, the unknown plane parameter $h$ 
cancels. It sets the absolute scale and cannot be inferred from a single image.  Furthermore, the scaling function is computationally tractable and, 
as for the exponential scaling discussed above, the scaled distance function is available in closed form,
\begin{align}\label{eq:scaled distances}
d_c(x^i, & x^j) =  d(x^i, x^j) \ \times \ \gamma(\delta,h,f\big)^{\frac{1}{2}}\\
& \times \Big|\frac{2h\sqrt{f}}{\langle \delta, X^i\hspace{-.1cm}-\hspace{-.1cm}X^j\rangle}
\left(\frac{1}{\langle\delta, -X^i\rangle^{\frac{1}{2}}} - 
  \frac{1}{\langle \delta,-X^j\rangle^{\frac{1}{2}}}\right)\Big|, \nonumber 
\end{align}
provided that the basic requirement $\langle \delta, X ^i \rangle < 0$ is fulfilled for all $i = 1,\ldots, n$.
Examples of scaled distances are given in Fig.\ref{fig:our scaling}.  When compared with 
Fig.~\ref{fig:exponential scaling}, we see that  the perspective scaling in 
\eqref{eq:model intensity} results in similar distance scaling as the exponential scaling 
while it also provides a coherent description of the perspective foreshortening.

For a given image, we assume that the focal length $f$ is known. It remains to estimate the parameters 
$(\beta, \eta_1, \eta_2)$ of the intensity function in \eqref{eq:model intensity} based on the estimated point 
pattern $\Psi$. The desired 3D image information, the slant and the tilt of the surface, may then be characterized 
by the scaling parameter estimates $\hat{\eta}_1$ and $\hat{\eta}_2$.  The parameter estimation is performed 
by maximizing the composite likelihood, see e.g. \cite{Moeller2010}, that takes the form
\begin{equation}\label{eq:composite likelihood} 
L(\Psi|\beta, \eta_1, \eta_2) \propto \exp(-\beta |D|) \ \beta^n \prod_{i = 1}^n c_\eta^{-2} (x^i).
\end{equation}
The maximum composite likelihood estimate for $\beta$ is $\hat{\beta} = n/|D|$. For the remaining two 
parameters--the parameters of interest in our setting--we maximize the function
\begin{align}\label{eq:composite log-likelihood}
l(\Psi|& \hat{\beta}, \eta_1, \eta_2) \\
& = n \log\Big(\frac{n}{|D|}-1\Big) +  \sum_{i = 1}^n \log(c_\eta^{-2} (x^i)). \nonumber
\end{align}

\section{Results}\label{sec: results}

\begin{figure} 
\centering
\subfigure[$\delta\hspace{-.1cm}=\hspace{-.1cm}(\frac{1}{\sqrt{2}}, 0, \frac{1}{\sqrt{2}})^\top$]{\includegraphics[width=3.5cm, height=3.5cm]{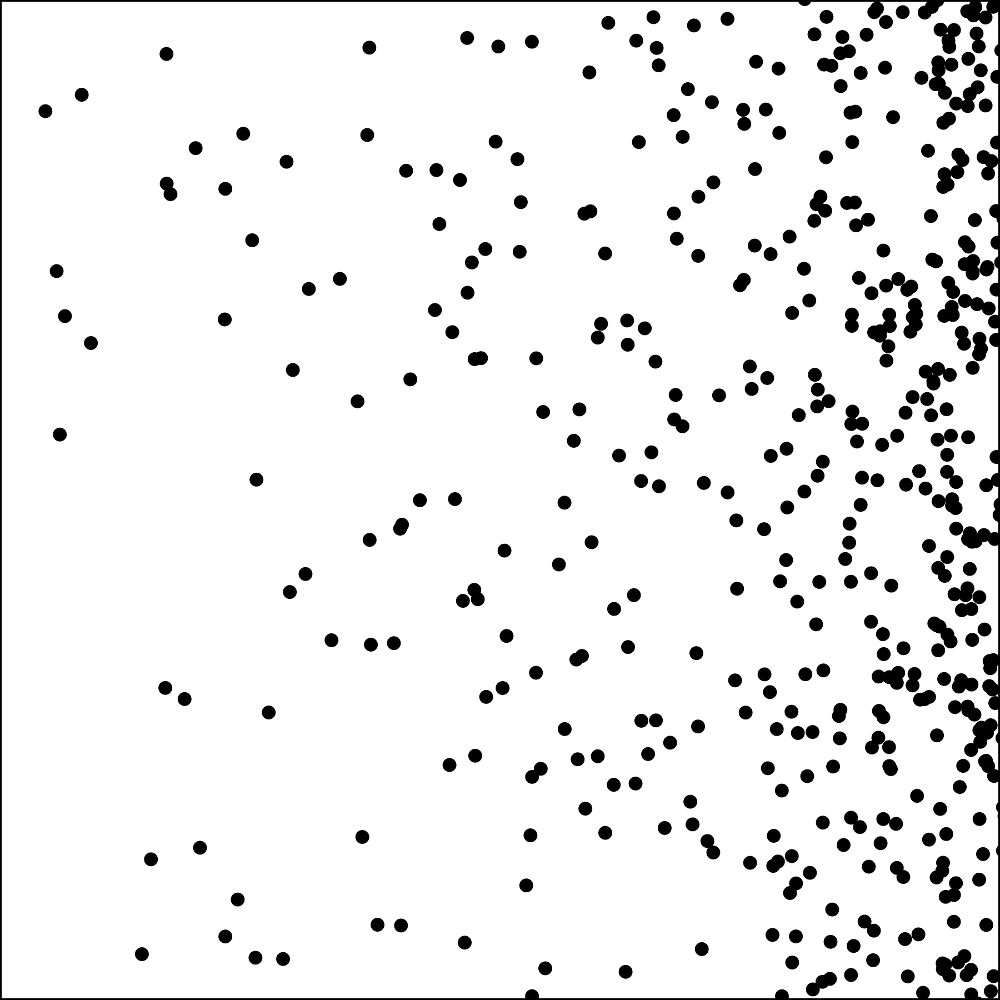}}
\hspace{0.5cm}
\subfigure[$\delta\hspace{-.1cm}=\hspace{-.1cm}(\frac{1}{2\sqrt{2}},\frac{1}{2\sqrt{2}},\frac{\sqrt{3}}{2})^\top$]{\includegraphics[width=3.5cm, height=3.5cm]{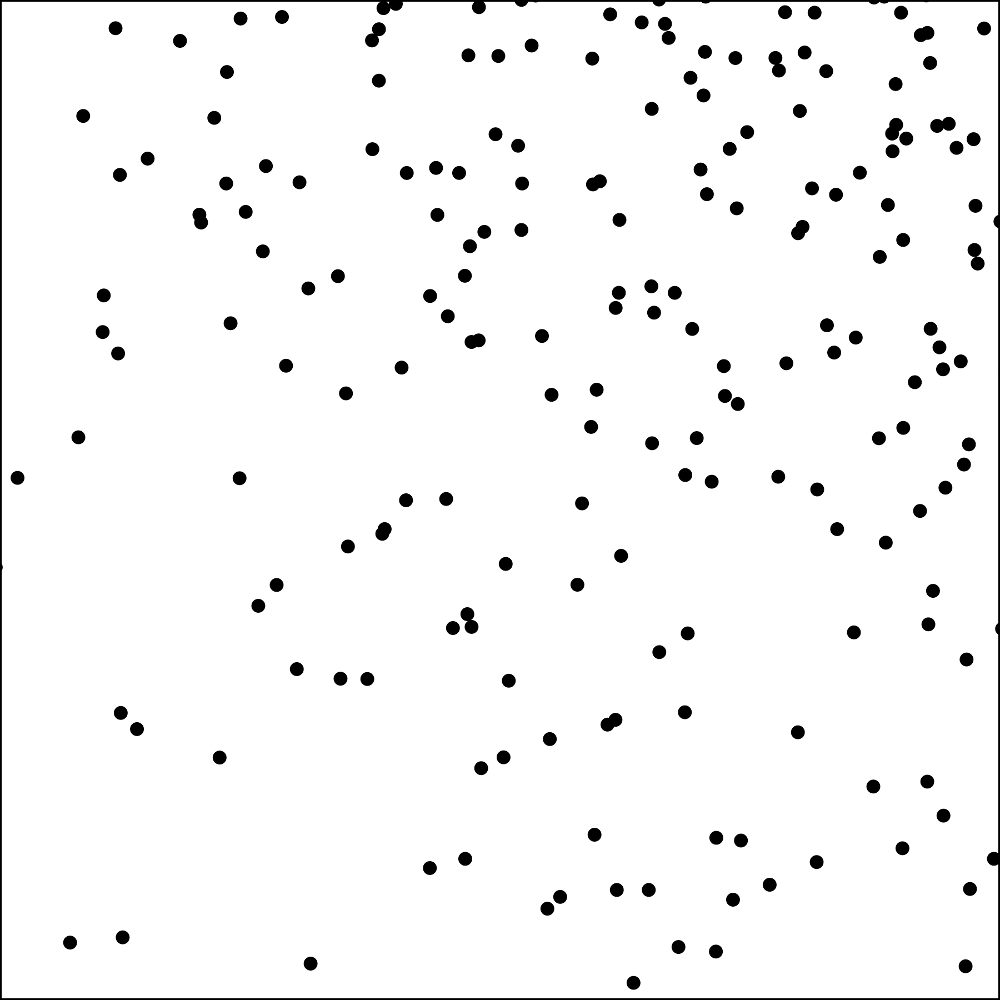}}
\caption{Simulated Poisson point patterns with 3D shape given by the outer normals in the subfigure captions. The internal parameters correspond to the settings in Fig.~\ref{fig:regular points} and Fig.~\ref{fig:our scaling}.} \label{fig:random points}
\end{figure}

\noindent
We first present the results of a simulation study where we analyse sets of 3D point coordinates sampled from either 
a perfectly regular pattern or a homogeneous Poisson processes and subsequently projected onto the 2D-plane 
$D=[-1/2, 1/2]\times[-1/2, 1/2]$, see Fig.~\ref{fig:regular points} and Fig. \ref{fig:random points}. 

We estimate the scaling parameters associated with the synthetic patterns via the composite likelihood in 
\eqref{eq:composite log-likelihood}.  The true parameter values and the corresponding estimates are given in 
Table~\ref{tab:simulation results}.  While the estimation procedure is able to reconstruct the true values with a 
resonable accuracy, the results are slightly better for the regular patterns than for the random patterns. These 
results are representative for several further such examples (results not shown), and we conclude that the 
composite likelihood is able to identify the scaling parameters of the perspective scaling function irrespective of the 
second order structure of the point pattern. 

\begin{table}[ht!]
\centering
\caption{True angles and composite likelihood estimates for the surface normals of the simulated point patterns 
in Figures~\ref{fig:regular points} and \ref{fig:random points}. Regular pattern type refers to the 
images in Figure~\ref{fig:regular points} and Poisson type to the images in 
Figure~\ref{fig:random points}.}\label{tab:simulation results}
\begin{tabular}{lcc}
\toprule
Pattern type &$(\eta_1, \eta_2)$&$(\hat\eta_1, \hat\eta_2)$\\
\midrule
Regular &$(45^{\circ}, 0^{\circ})$ & $(45.5^{\circ}, 0.0^{\circ})$\\
Poisson &$(45^{\circ}, 0^{\circ})$&$(46.2^{\circ}, 0.7^{\circ})$\\
Regular &$(30^{\circ}, 45^{\circ})$ & $(29.9^{\circ}, 45.7^{\circ})$\\
Poisson &$(30^{\circ}, 45^{\circ})$&$(26.2^{\circ}, 45.5^{\circ})$\\
\bottomrule
\end{tabular}
\end{table}

\begin{figure}
\centering
\subfigure[Tiling A]{ 
 \includegraphics[width=.225\textwidth]{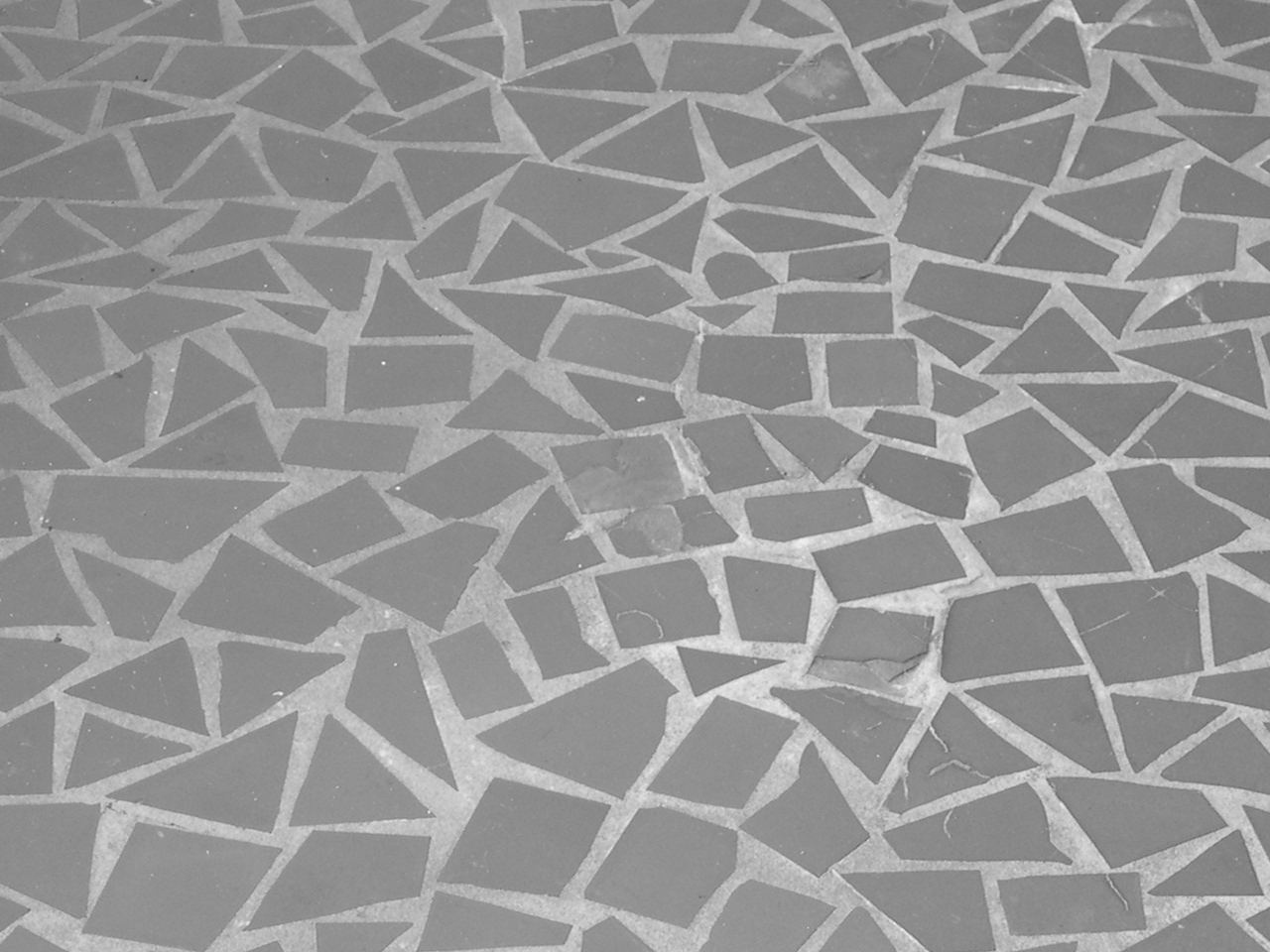}\hspace{.85cm}
 \includegraphics[width=.225\textwidth, height=3.1cm]{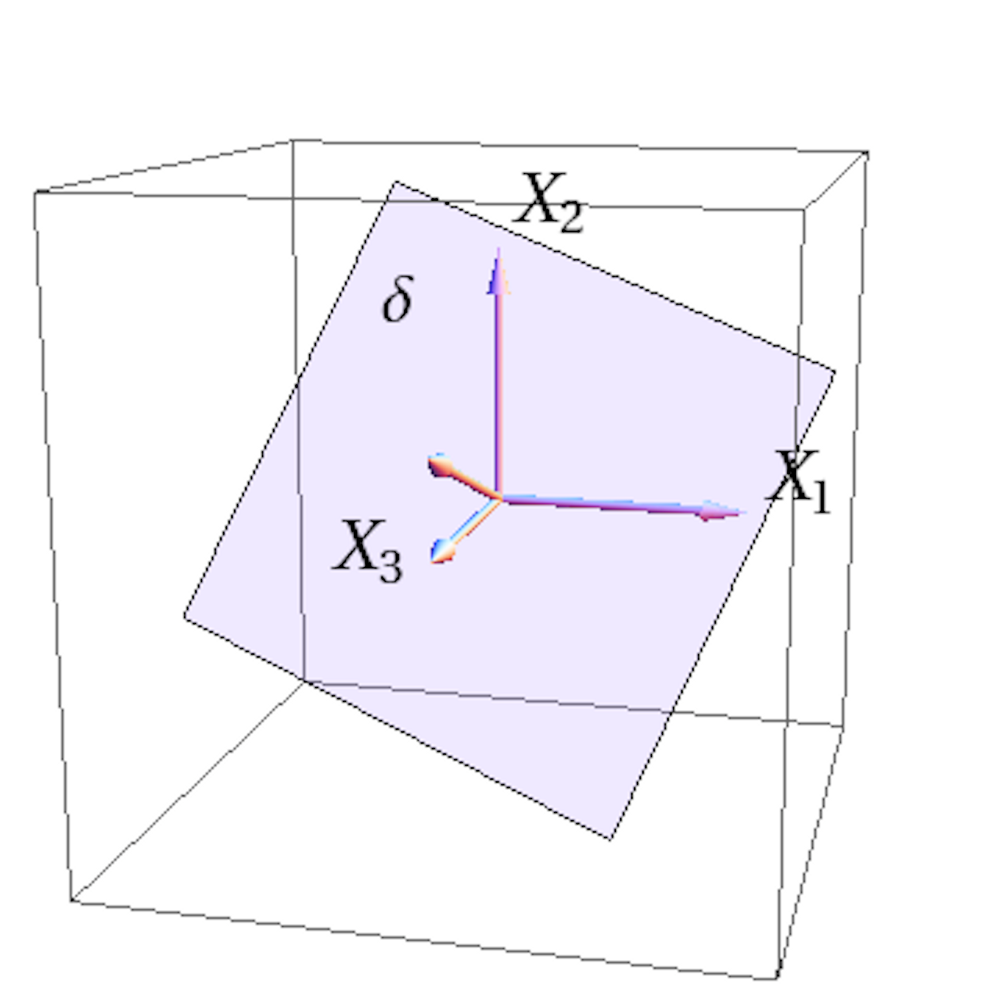}
 }
\subfigure[Tiling B]{ 
 \includegraphics[width=.225\textwidth]{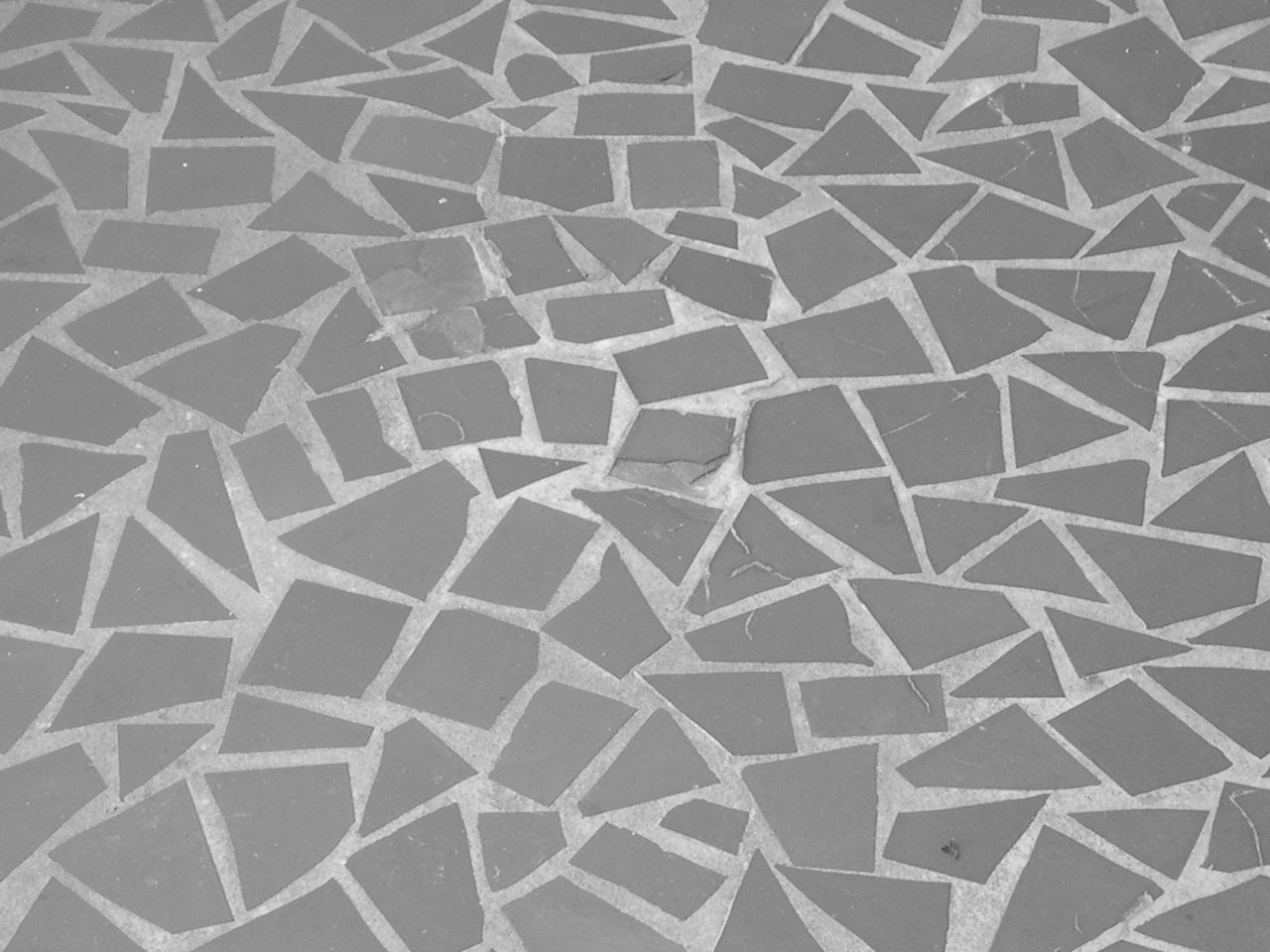}\hspace{.85cm}
 \includegraphics[width=.225\textwidth, height=3.1cm]{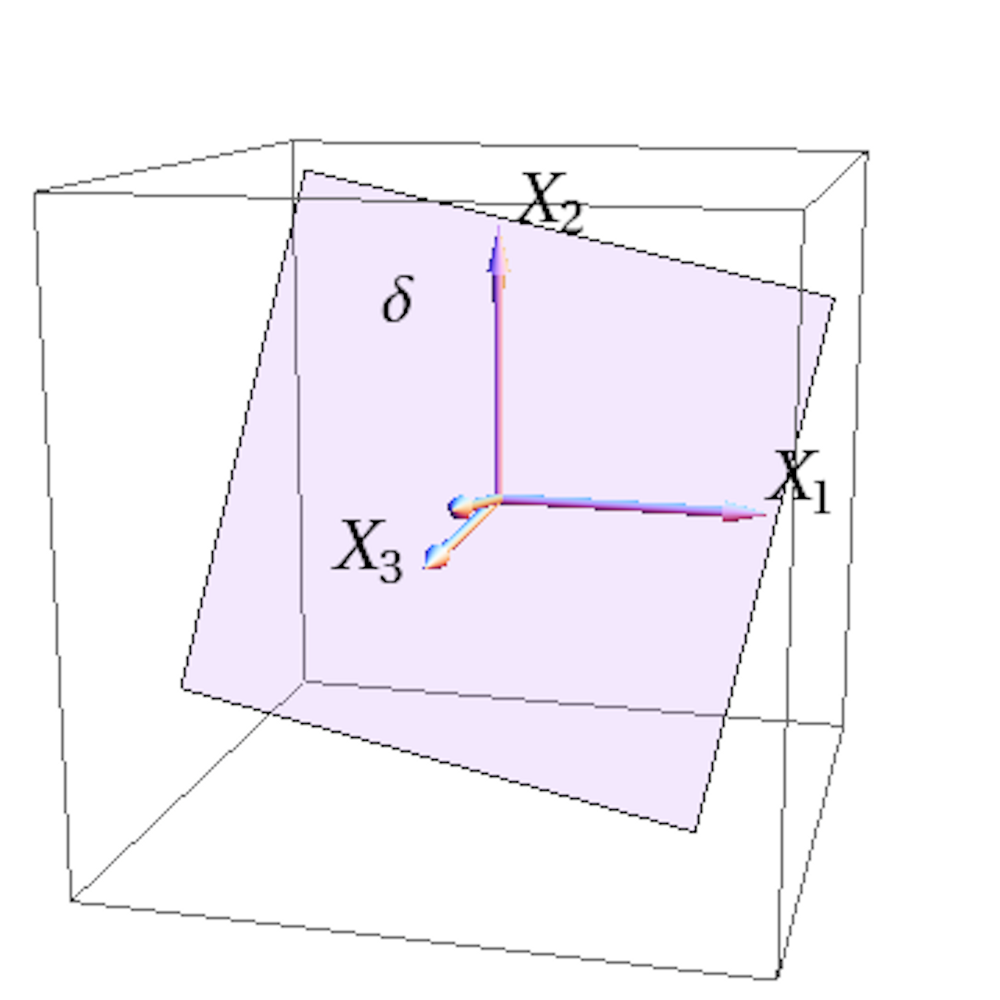}
 }
\subfigure[Bricks]{ 
  \includegraphics[width=.225\textwidth]{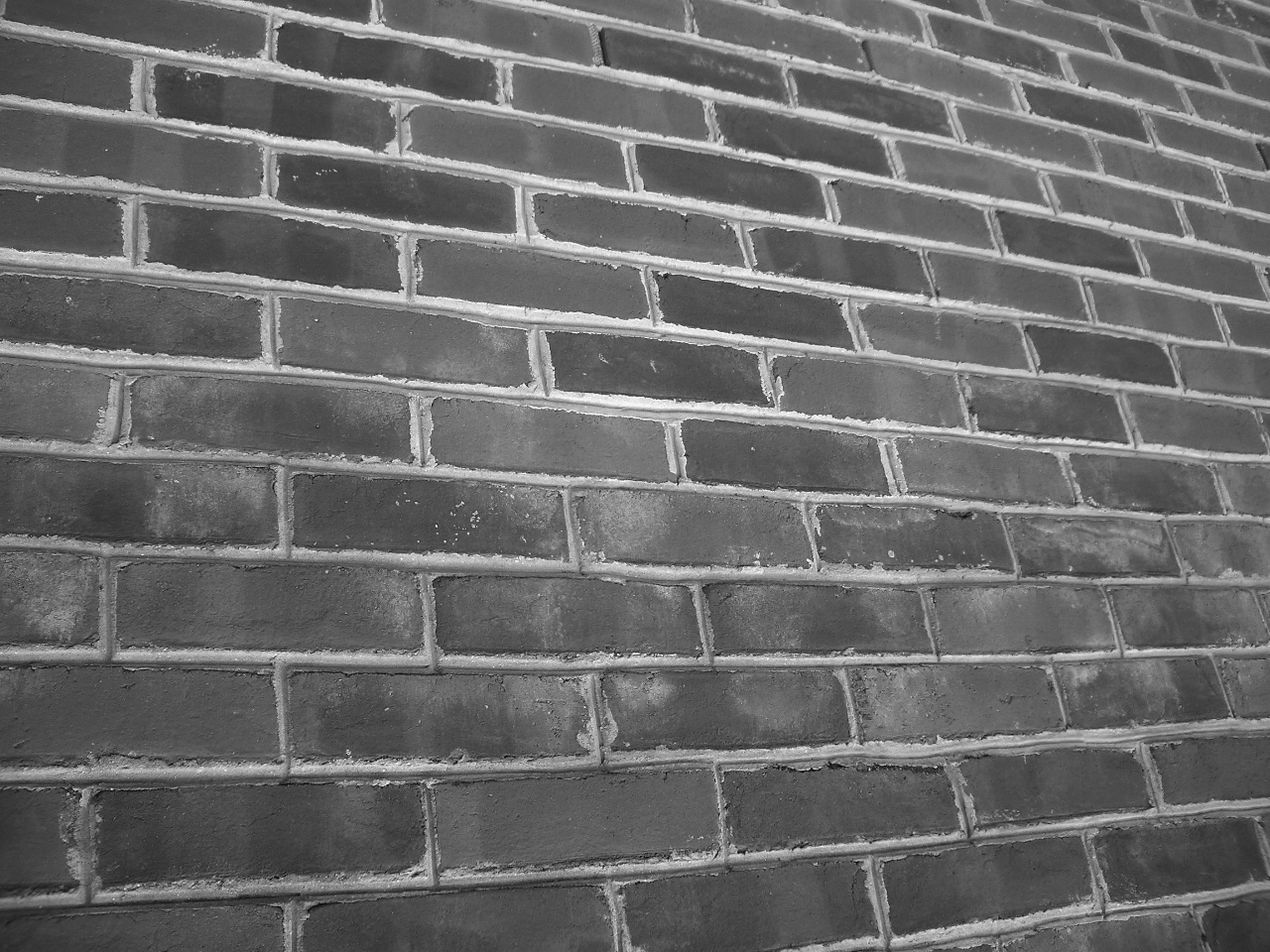}\hspace{.85cm}
  \includegraphics[width=.225\textwidth, height=3.1cm]{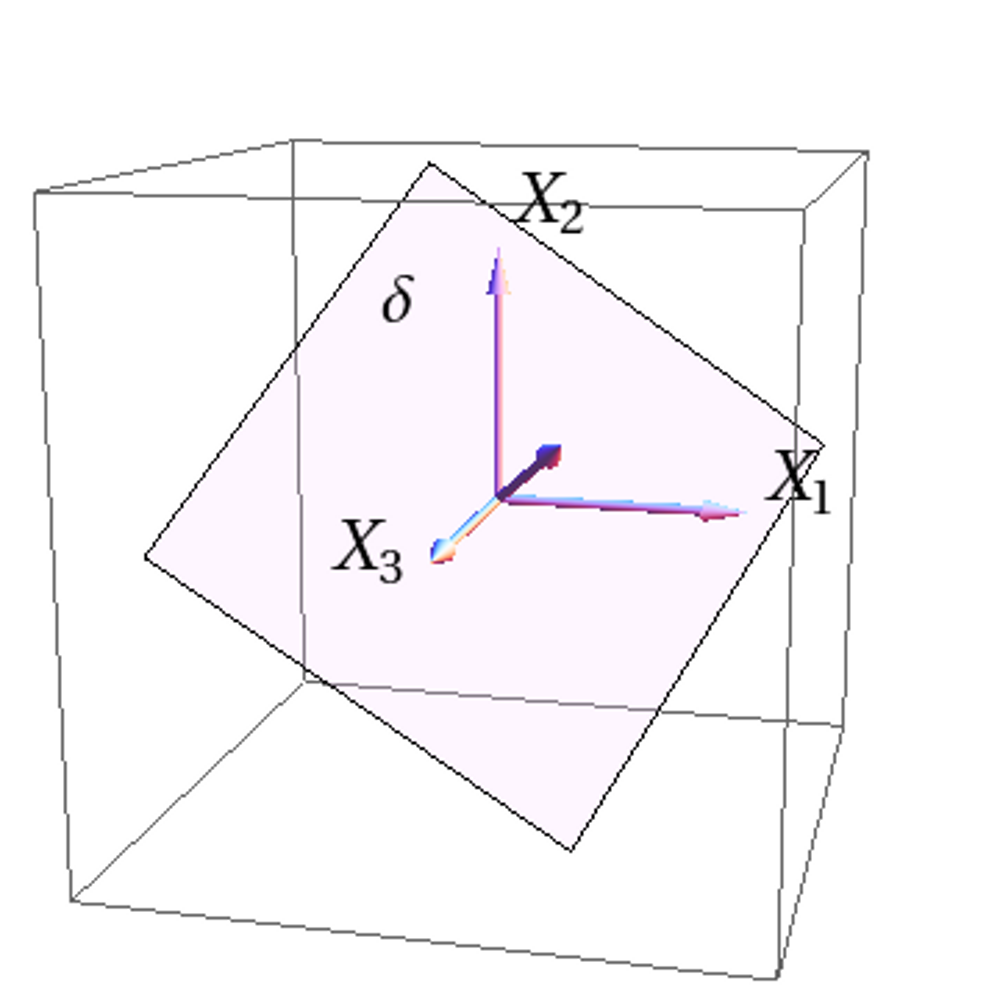}
}
\caption{Original natural scenes  (left) and the estimated 3D orientation towards the camera (right). The field of view is assumed to be driven by a wide angle setting of $\phi_c = 54^{\circ}$.} 
\label{fig:image orientation}
\end{figure}

\begin{figure} 
\centering
\subfigure[Tiling A]{ 
 \includegraphics[width=.45\textwidth, height=6cm]{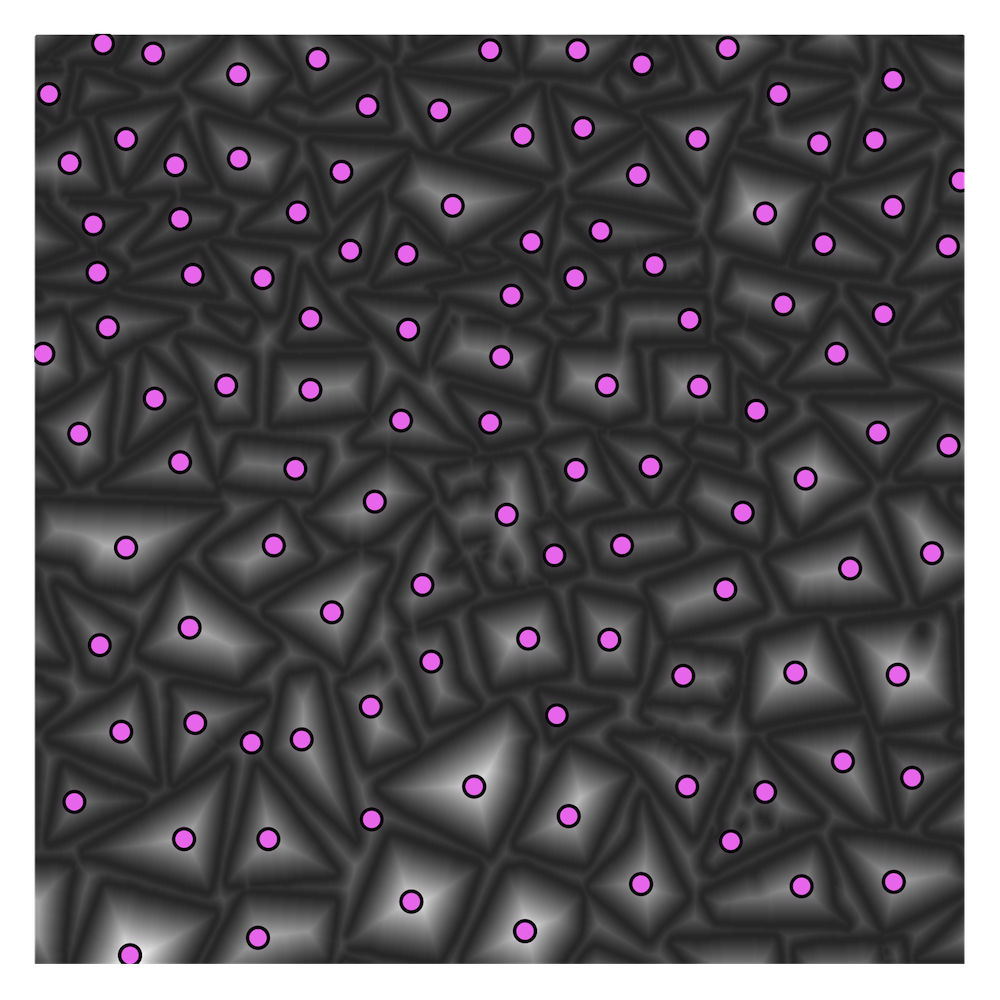}
 }\hspace{.05cm}
\subfigure[Tiling B]{ 
 \includegraphics[width=.45\textwidth, height=6cm]{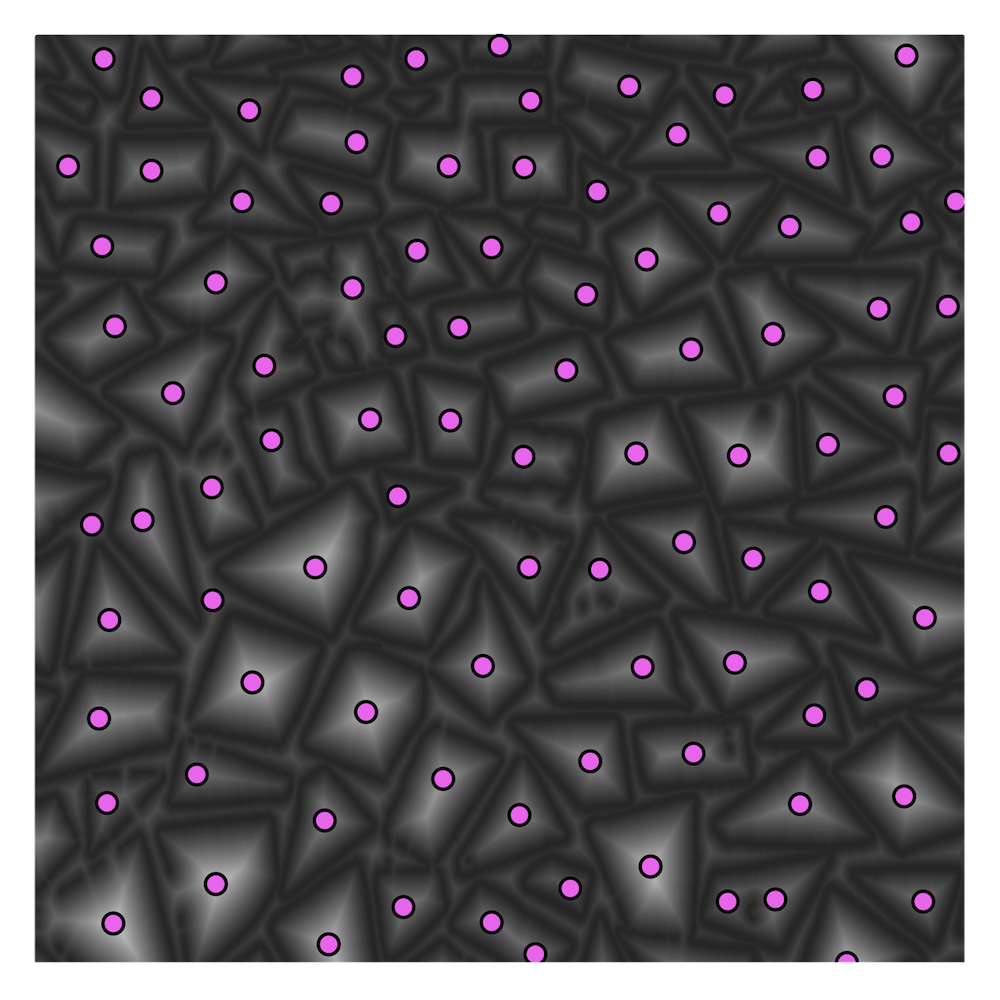}
 }
\subfigure[Bricks]{ 
  \includegraphics[width=.45\textwidth, height=6cm]{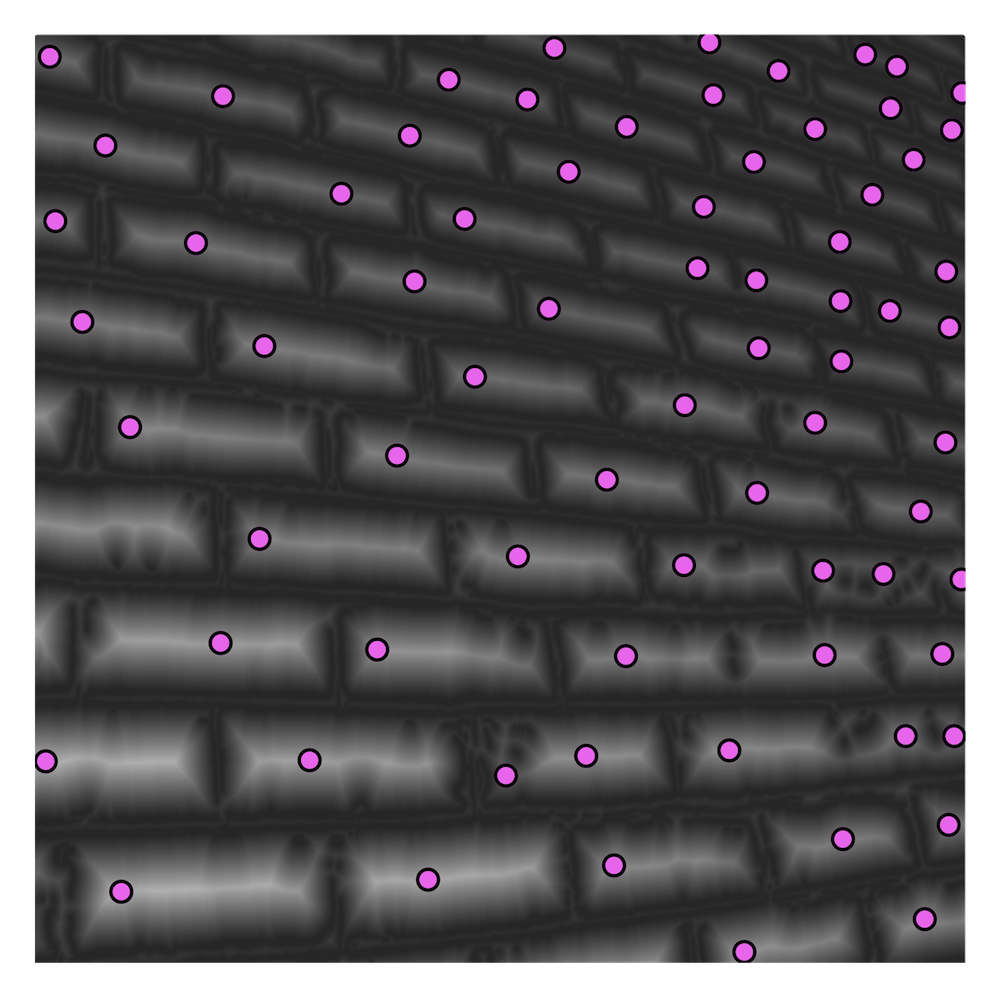}
}
\caption{Estimated probability maps and point configurations for the natural scenes in Fig.~\ref{fig:image orientation}.}
\label{fig:image points}
\end{figure}

For the analysis of real natural scenes, we apply our methodology to the set of tiling and brick images shown in Fig.~\ref{fig:image orientation}.  The original images are of size $1280\times960$ pixels and during the preprocessing they are downsided to $1066\times846$ pixels in order to eliminate boundary effects in the point detection.  The probability maps and the resulting point patterns are shown in Fig.~\ref{fig:image points}.  We have here applied neighbourhoods of sixe $75 \times 75$ pixels for the tiling scenes and $55 \times 55$ pixels for the bricks scene, with a threshold of $k_2 = 0.25$ for the neighbourhood relation in all cases.  The point detection is very robust in the selection of threshold value and threshold values from $0.15$ to $0.5$ have limited effects on the results.  It is somewhat more sensitive to changes in the neighbourhood size; for the tiling images neighbourhoods from $55 \times 55$ to $95 \times 95$ result in similar  scaling parameter estimates while for the bricks image, slightly smaller neighbourhoods seem to be needed. 

For deriving the information on camera positioning and angle from the point configurations in Fig. ~\ref{fig:image points}, we project the point process realizations onto an observation window $D$ of dimension $[-0.69, 0.69]\times[-0.50, 0.50]$.  We further assume that the field of view corresponds to a standard wide angle setting of $\phi_c = 54^{\circ}$ and hence take $f=0.98$ as a basis,  the same settings as we applied in the simulation examples above.  The resulting scaling parameter estimates are listed in Table \ref{tab:data results} and the 3D orientation of the camera toward the textures is illustrated in Fig. \ref{fig:image orientation}. 

\begin{table}[ht!]
\centering
\caption{Perspective scaling parameter estimates for the natural scenes in Fig. \ref{fig:image orientation}.}\label{tab:data results}
\vspace{.2cm}
\begin{tabular}{lc}
\toprule
Texture type & $(\hat\eta_1, \hat\eta_2)$ \\
\midrule
(a) Tiling A&$(22.1^{\circ}, 94.7^{\circ})$\\
(b) Tiling B&$(12.2^{\circ}, 65.9^{\circ})$\\
(c) Bricks& $(36.0^{\circ}, 44.1^{\circ})$ \\
\bottomrule
\end{tabular}
\end{table}

\section{Discussion}\label{sec: discussion}

\noindent
This paper introduces a framework for extracting 3D information from a textured 2D image building on the recently 
developed locally scaled point processes \citep{Hahn}. The perspective scaling function quantifies perspective 
foreshortening and the resulting inhomogeneity of the texture.  The framework is quite flexible regarding assumptions on 
the texture composition in that it only requires the texture elements to be close to convex in shape and it successfully 
extracts useful information related to camera orientation.  

The separation of image preprocessing and point detection on one hand and the estimation procedure for the scaling 
parameters on the other hand offers great flexibility. We believe that the locally scaled point process framework can
be applied in more general settings to analyse point patterns in images, for instance, as a new additional inference step 
in the texture detection algorithms discussed in \citet{Lafarge2010} and references therein.  Due to the low
computational budget of our framework, it also seems feasible to combine it with image segmentation where 
3D information is needed for several segments within an image, each of which might be covered with a different type 
of texture elements.    

There are further considerable avenues for development. One area for future development is to build a large hierarchical framework where the three inference steps, the image preprocessing, the point detection and the parameter estimation, are joined in an iterative fashion.  
A fully Bayesian inference framework along the lines of the work of \citet{Rajala2012} could also be an alternative to the composite likelihood estimation performed here. Future work will concentrate on embellishing our inference framework. 

\section{Acknowledgments}

We thank Ute Hahn for sharing her expertise.  This work has been supported by the German Science Foundation (DFG), grant RTG 1653. The work of Thordis L. Thorarinsdottir and Alex Lenkoski was further supported by Statistics for Innovation, {\em sfi}$^2$, in Oslo.

\section{Appendix}

In our data analysis, we assume that the image domain is normalized such that $D = [-a/2,a/2] \times [-1/2,1/2]$. 
More generally, the image domain could be of the form $D = [a_1,a_2] \times [b_1,b_2]$ for some $a_1,a_2,b_1,b_2 \in \real$ with $a_1 < a_2$ and $b_1 < b_2$.  In this case, the condition of conservation of the total area in \eqref{eq:conservation area} becomes 
\begin{equation}
 |D| = (a_2-a_1)(b_2-b_1) = \int_{D} \gamma (\delta, d, f) dX_{P}.
\end{equation}
It follows that  
\begin{align}\label{eq:general normalizing} 
\gamma (\delta, h,f) \, = \, & \frac{2}{h^2 f}  (-(a_1+a_2)\delta_1 - (b_1+b_2)\delta_2 +f\delta_3)^{-1}\nonumber\\
& \quad \times (a_1 \delta_{1} + b_1 \delta_{2} - f\delta_3) \nonumber\\
& \quad \times (a_1 \delta_{1} + b_2 \delta_{2} - f\delta_3)\\
& \quad \times (a_2 \delta_{1} + b_1 \delta_{2} - f\delta_3) \nonumber\\
& \quad \times (a_2 \delta_{1} + b_2 \delta_{2} - f\delta_3) \nonumber \ . \nonumber
\end{align}

\end{document}